\documentclass{optica-article}

\journal{oe}

\articletype{Research Article}

\usepackage{lineno}
\usepackage{amsmath}
\usepackage[normalem]{ulem}
\usepackage{float}
\usepackage{tabularx}
\usepackage{lscape}
\usepackage{float}
\linenumbers

\begin{document}

\title{A Computer Vision Based Optical Method for Measuring Fluid Level in Cell Culture Plates}

\author{Pierre V. Baudin\authormark{1,2,*} Mircea Teodorescu\authormark{1,2}}

\address{\authormark{1}Department of Electrical and Computer Engineering,   University of California Santa Cruz \\ Santa Cruz, CA 95064, USA\\
\authormark{2}Genomics Institute, University of California Santa Cruz \\ Santa Cruz, CA 95064, USA\\
}

\email{\authormark{*}pvbaudin@ucsc.edu} 




\begin{abstract}
For a transparent well with a known volume capacity, changes in fluid level result in predictable changes in magnification of an overhead light source. For a given well size and fluid, the relationship between volume and magnification can be calculated if the fluid’s index of refraction is known or in a naive fashion with a calibration procedure. Light source magnification can be measured through a camera and processed using computer vision contour analysis with OpenCV. This principle was applied in the design of a 3D printable sensing device using a raspberry pi zero and a camera
\end{abstract}

\section{Introduction}\label{sec:intro}


The automation of cell culture procedures has the potential to greatly increase the throughput and consistency of cell culture based experiments \cite{doulgkeroglou_automation_2020}. Manipulation of fluids of is a key feature for any automated culture platform. Systems must have the ability to deliver or move precise quantities of fluids. To automate movement of fluids, some systems use a "lab-on-a-chip" approach \cite{figeys_lab---chip_2000}, utilizing microfluidic channels to transport fluids. Other systems use robot manipulators to deliver measured quantities of fluids with micro-pipettes \cite{doulgkeroglou_automation_2020, fleischer_analytical_2018, steffens_versatile_2017}. Regardless of mechanism behind the delivery of fluids, being able to detect the presence and volume of fluids within provides key error detection and correction functionality that open loop methods lack.

Surface level measurement within a container of known volume is a simple way to determine fluiod volume. This can be done with electrically active probes dipped into a fluid being measured \cite{singh_review_2019}. These probes can be effective but direct contact with the fluid can be problematic for cell culture applications wherein any potential contamination vector increases the risk of failure \cite{lincoln_chapter_1998}. Non-contact sensing is preferable for sensitive applications. Optical fluid sensing methods take advantage of how fluids interact with light. Several optical approaches for measuring fluids in culture plates exist. One method uses computer vision with well placed light sources and sensors placed at specific angles to measure ray deflection \cite{jain_total-internal-reflection_2021}. This method uses reflectance of the fluid surface, which makes it usable no matter the solid mass contents of the well. However, the angular requirements of the sensor electronics makes tight packing of independent sensing elements infeasible, making parallel measurements within a plate complex or impossible. Another published method involves imaging the distortions in a printed grid to visualize changes in refraction related to fluid level \cite{litt_visualization_1989}. While this approach can be used in parallel on many wells at once, it is highly affected by occlusion resulting from material in the plate and can be affected by ambient lighting conditions.



Here we propose a method for non-contact optical detection of fluid level using a CMOS camera sensor and an LED. We detail the principles at work and show a 3D printed device that can be used to perform measurements in 24 well culture plates. Similar to the approach shown in \cite{litt_visualization_1989}, this method relies on the lens-like properties of fluids and the resulting changes of an image viewed through them. Instead of viewing a grid, we detect the apparent size of an overhead light source. Changes in the apparent distance of the light source correlate to changes in fluid level. A bright overhead light source can shine through some samples, making this approach usable in scenarios with solid occlusion that would disrupt the viewing of a grid. Additionally, the electronics required are set up directly inline with the sample wells, allowing electronics to be packed into a grid allowing parallel sensing of many wells at once. 

\section{Methods}\label{sec:methods}
\subsection{Operating Principle}
\label{subsec:principle}
\begin{figure}[H]
    \centering
    \includegraphics[width = \columnwidth]{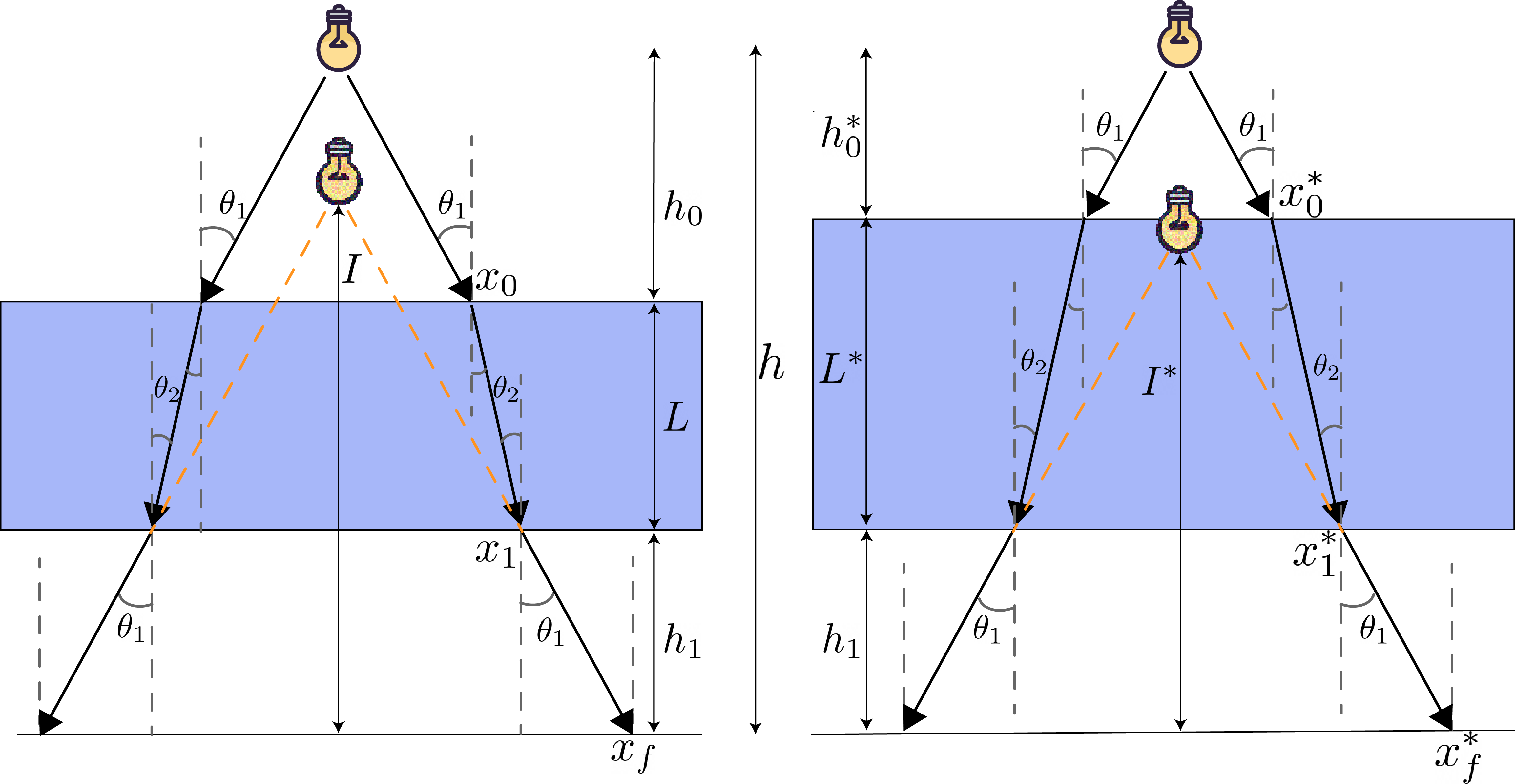}
    \caption{\textbf{Apparent distance of the light source changes based on fluid depth for fluids with index of refraction greater than air} (relationship is inverted if index is less than air). Refraction angles can be determined with Snell's law with known indices of refraction of the 2 media ($n_1 sin(\theta_1) = n_2 sin(\theta_2)$). In the case of second transition back into the original medium, the relationship $\theta_3 = \theta_1$ can be derived, hence the labeling of the third angle as $\theta_1$ in this diagram. $L$ = fluid depth, $I$ = apparant distance of object. $h_0$ = distance between object and top of fluid, $h_1$ = distance between bottom of fluid and measurement plane.}
    \label{fig:principle}
\end{figure}

The measurement principle behind this system relies on the lensing effects of fluids with different indices of refraction. In figure \ref{fig:principle} the diagrams represent how the apparent distance of the light source changes based on fluid depth. In the case of a fluid with a index of refraction greater than air (like water), the light source will appear closer as the fluid level increases. \\

To derive a function relating the fluid depth to the apparent distance of the light source, we define the following constants.
$$
\begin{aligned}
& C_0 = h - h_1 \\
& C_1 = \tan \theta_{1} \\
& C_2 = \tan \theta_{2}
\end{aligned}
$$
These can be considered constants because our light source sends rays in all directions below it, so any angle $\theta_1$ can be chosen so long as it results in a ray that intersects the fluid, and from this chosen angle $\theta_2$ can be derived via Snell's Law. In terms of these constants the resulting transfer function is:
$$
I=\frac{L(-C_1 + C_2) + h_1C_1 + C_0C_1}{C_0}
$$
This function relates the apparent distance of the image $I$ to the fluid level $L$ and is a first degree polynomial. The full derivation can be found in section 1 of the supplemental materials. This function is a useful approximation that ignores several factors including refraction caused by the plate material and refraction from the meniscus geometry of the fluid surface, our data show that accurate results can be obtained with this approximation in most scenarios. Meniscus has a substantial effect on measurement in two extremes, an almost empty well, and an almost full well. These meniscus effects are further explored in section \ref{sec:results}




The polynomial can be characterized analytically so long as we can determine how the image distance relates to size on the camera sensor. This can be determined by taking a picture of a ruled measurement calibration slide at several known distances. It can also be characterized by a simple calibration step, being that the relationship is linear, a 2 point calibration would theoretically be sufficient, but for greater accuracy a 5 point calibration would be more prudent. Once this polynomial is characterized, image data is sufficient for capturing fluid level

\subsection{Hardware}
\begin{figure}[H]
    \centering
    \includegraphics[width=\textwidth]{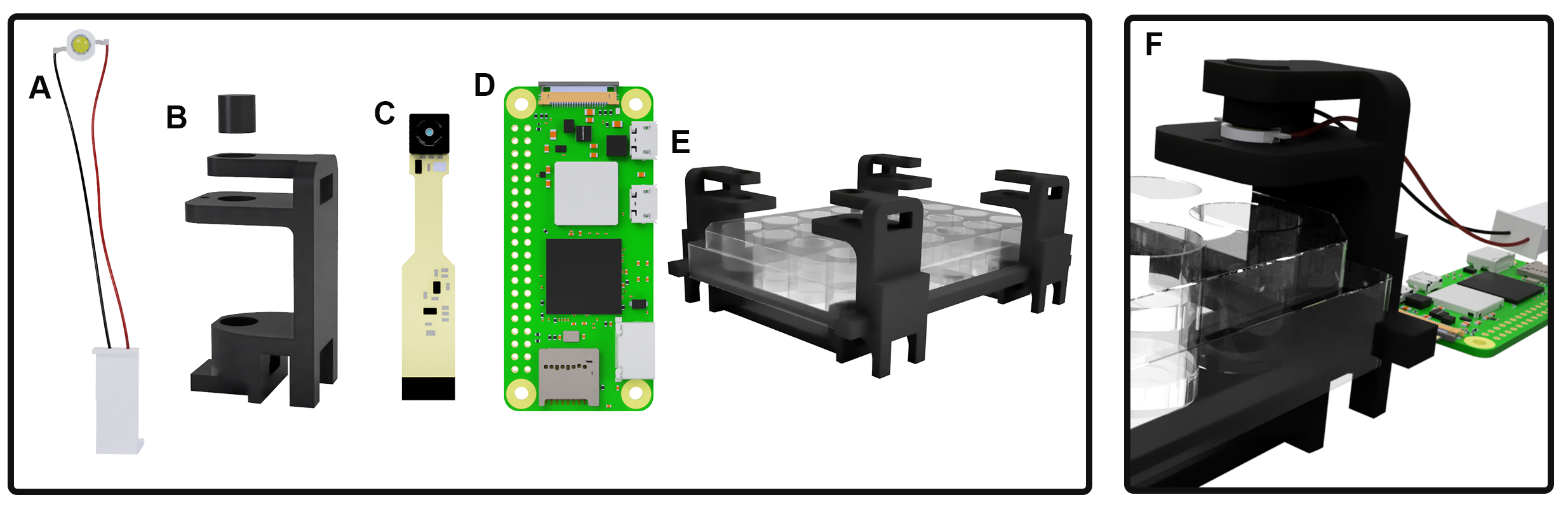}
    \caption{\textbf{3D renders of measurement rig} A: a 3W white LED, B: 3D printed enclosure with a plug to hold the light in place, C: Raspberry Pi Spy Camera, D: Raspberry Pi Zero W, E: rigid plate holder, F: experimental setup. Rendered using Fusion360, real setup shown in figure \ref{fig:device-with-plate} }
    \label{fig:exploded-view}
\end{figure}

\begin{figure}[H]
    \centering
    \includegraphics[width=\textwidth]{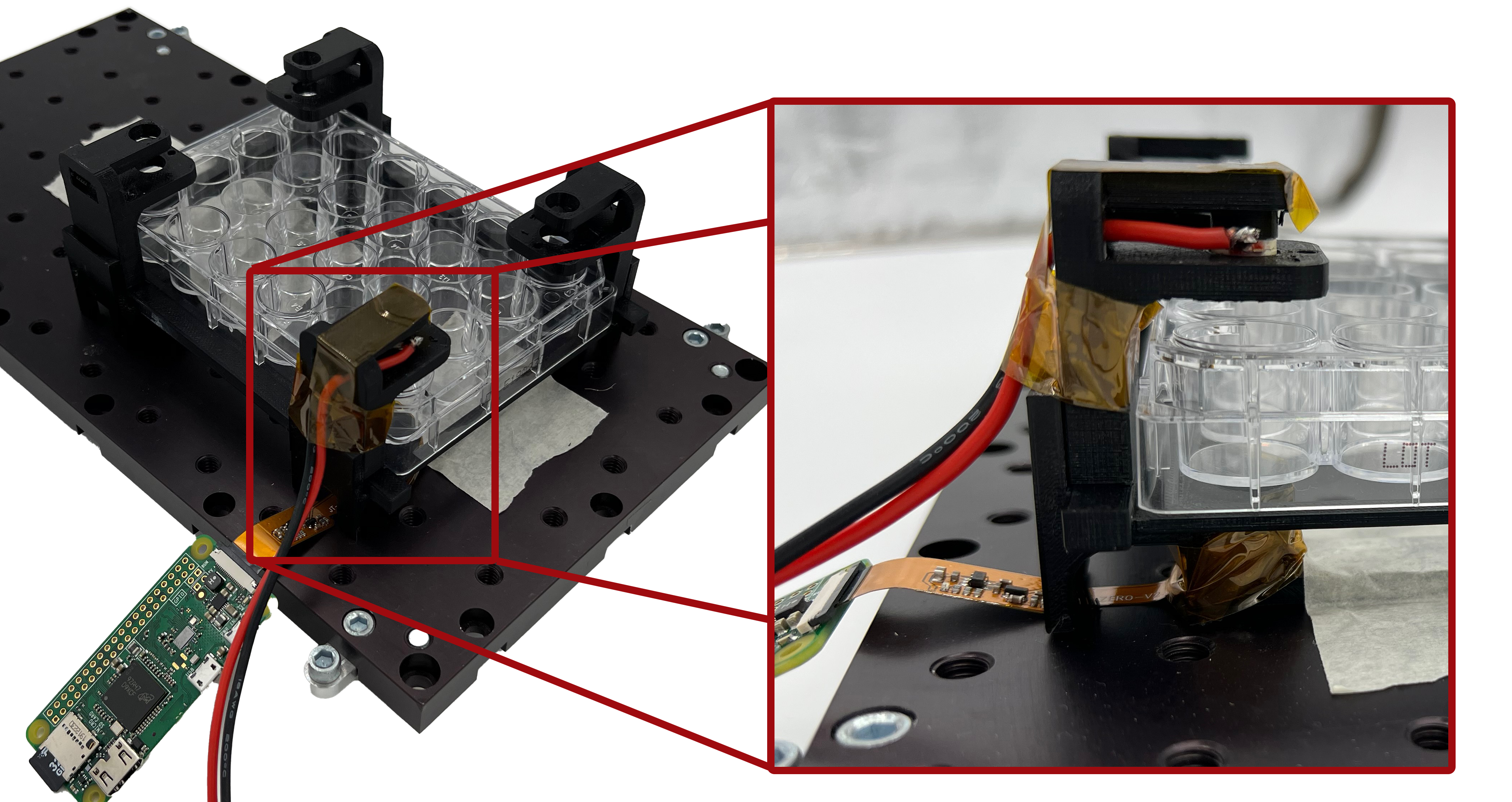}
    \caption{\textbf{Experimental setup with 24 Well plate} }
    \label{fig:device-with-plate}
\end{figure}

To test this approach, we designed a 3D printable rig that holds a raspberry pi camera and a white LED. Figure \ref{fig:exploded-view} shows 3D renders of the parts comprising the measurement device and figure \ref{fig:device-with-plate} shows the experimental setup.


\subsection{Analysis Pipeline}
Using openCV \cite{opencv_library}, a free and open-source machine vision toolkit, we can capture and measure the size of spot light image. This can be done on static images or on videos including live feeds. We do this by detecting the central contour in the image and fitting an ellipse to it. The result can be calculated fast and is robust to occlusion and noise. In order to do this On the raspberry pi zero we run a video stream using the open-source RPi Cam Web Interface \cite{noauthor_rpi-cam-web-interface_nodate}. The python code for this image analysis is lightweight and can be run internally on the raspberry pi or on an external system. For external system use, the code grabs frames from the MJPEG stream generated by the camera web interface.

\begin{figure}[H]
    \centering
    \includegraphics[width=\textwidth]{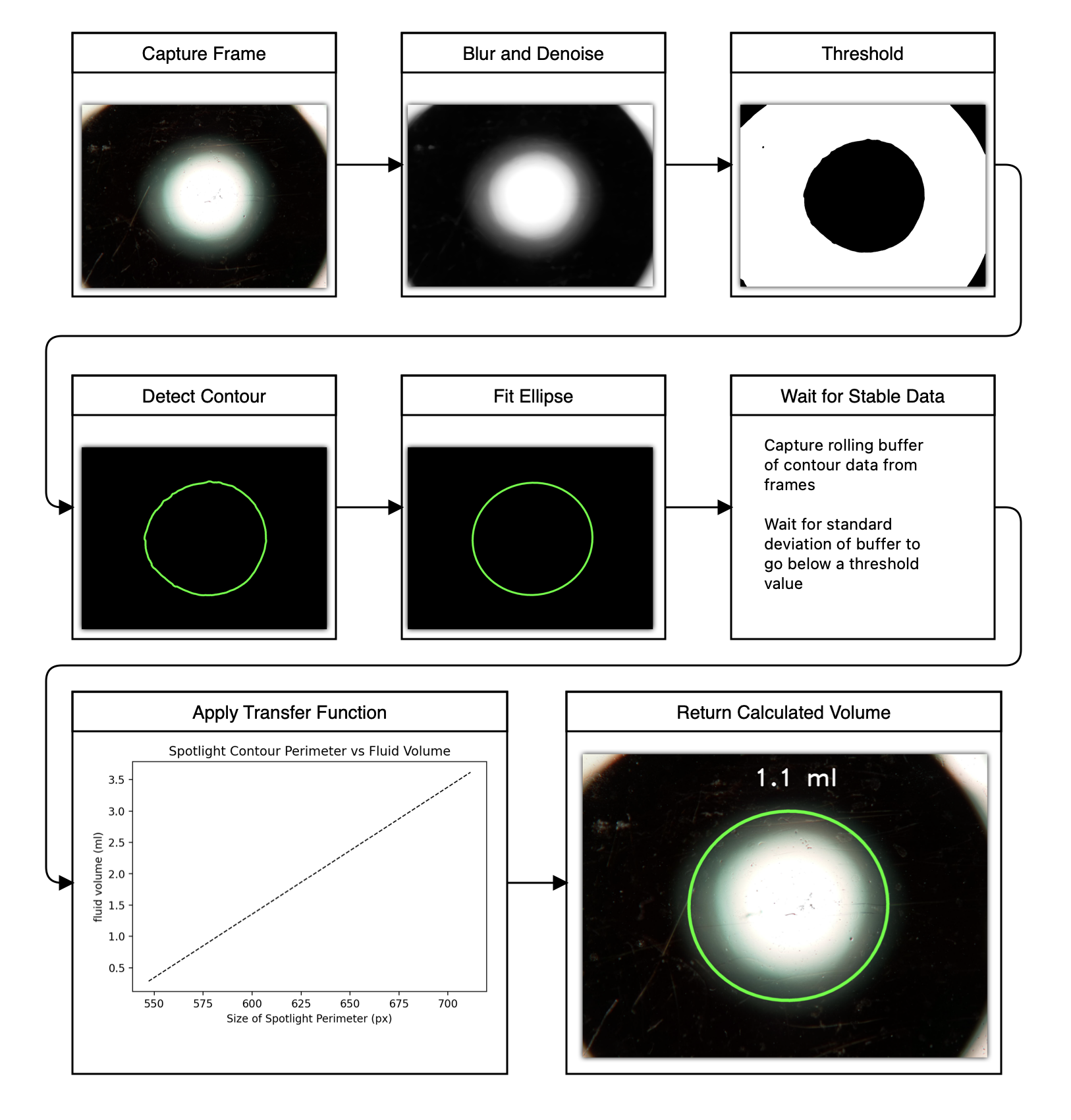}
    \caption{\textbf{Flowchart representing detection process pipeline}}
    \label{fig:detection-process}
\end{figure}

Figure \ref{fig:detection-process} details the process through which a data point is obtained. Once an ellipse is fit to the central contour of the image, we filter the output by accepting the average value of a rolling buffer when the standard deviation of that buffer falls below a threshold value. This is to compensate for the disturbance of the fluid surface when new fluid is added or if the plate is shaken. By only accepting reads when the reading has stabilized, we avoid wildly fluctuating erroneous readings any time the fluid is disturbed.

\section{Results}\label{sec:results}

\begin{figure}[H]
    \centering
    \includegraphics[width=\textwidth]{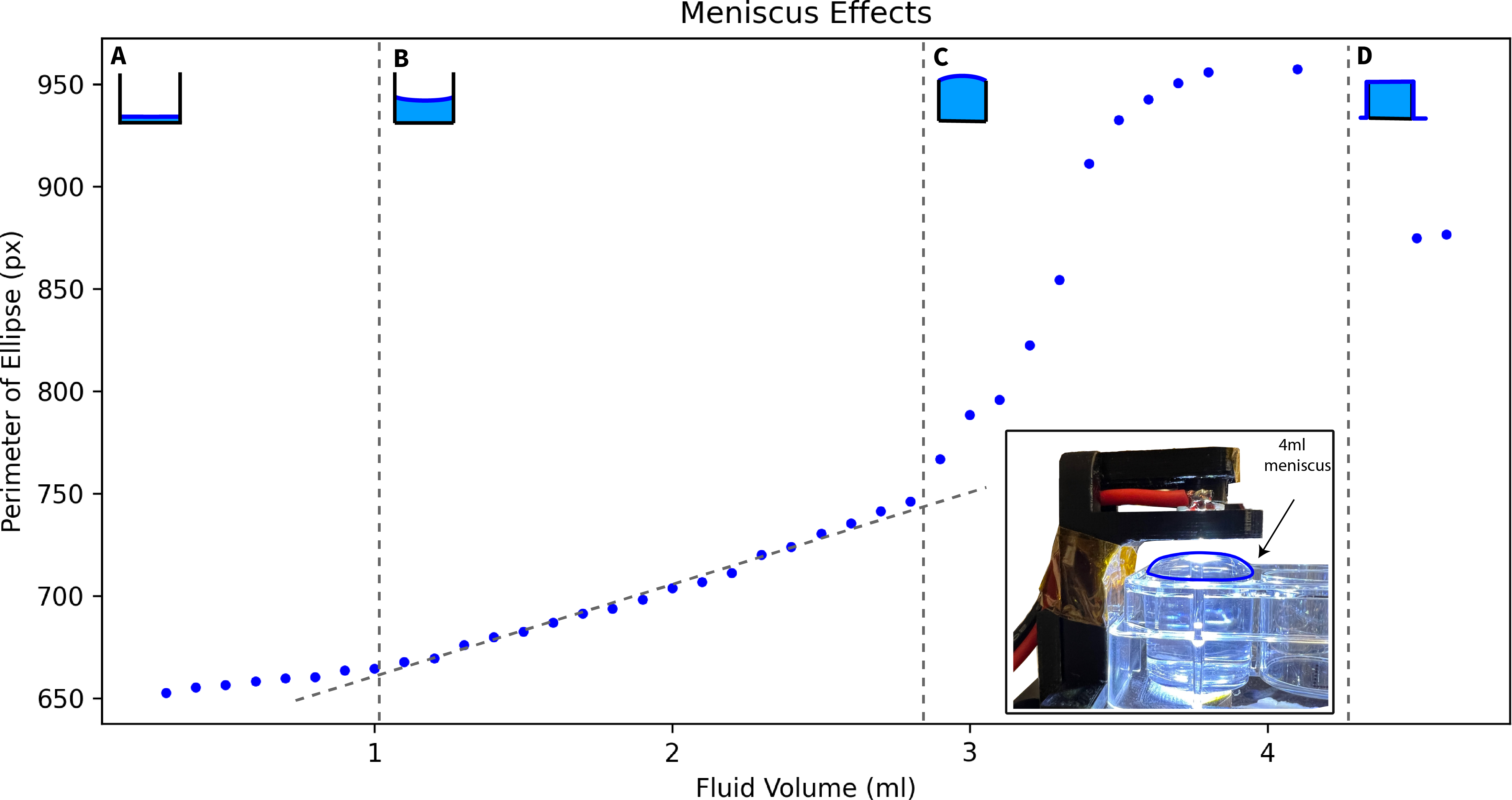}
    \caption{\textbf{Contour ellipse perimeter as a function of fluid volume.}
    Region A shows behavior before meniscus has reached stable geometry. Region B has stable meniscus and shows linear response as predicted in section \ref{subsec:principle}, Region C shows when the well begins to get full, the surface tension causes Meniscus Inversion. Region D is where fluid has spilled over the edge of the well, flattening the meniscus.}
    \label{fig:data-with-meniscus}
\end{figure}

The plot shown in figure \ref{fig:data-with-meniscus} shows the output of our sensing system for from the first possible reading to the overflowing of the well. Figure \ref{fig:Dry-Fill} shows why the first possible reading from a previously dry well occurs around 0.4ml. When the fluid level is very low in a well, the geometry of the meniscus cannot fully form. The center of the meniscus will rise more slowly before the full geometry of the meniscus is formed, this is because of the fluid volume held in the meniscus edges. Therefore, before the full formation of the meniscus, the overhead light image viewed from the center of the meniscus grows more slowly up until point A shown in figure \ref{fig:data-with-meniscus}. As the well starts to get full, a new distortion of the meniscus geometry occurs due to the surface tension of the fluid. The image grows substantially faster due to the magnifying properties of the convex meniscus, this phenomenon begins at point B. Finally, when the well overflows, the meniscus flattens resulting in a final level of unchanging magnification beginning at point C.

The region between points A and B are where this sensing process will be most generalizable, as the relationship between the spotlight image contour size and the fluid level is highly linear. 

\subsection{Calibration}
Deriving an effective transfer function for this system can be done by fitting a curve to a set of calibration points. Deciding on the best fit curve is a question of trade offs. The phenomenon we are measuring is linear within the range of volumes that have a stable meniscus geometry. A simple 2 point linear fit with points from the linear region, generates outputs with average error below 100$\mu$l within the linear range. Outside this range, the estimation error increases substantially. 

\begin{figure}[H]
    \centering
    \includegraphics[width=\textwidth]{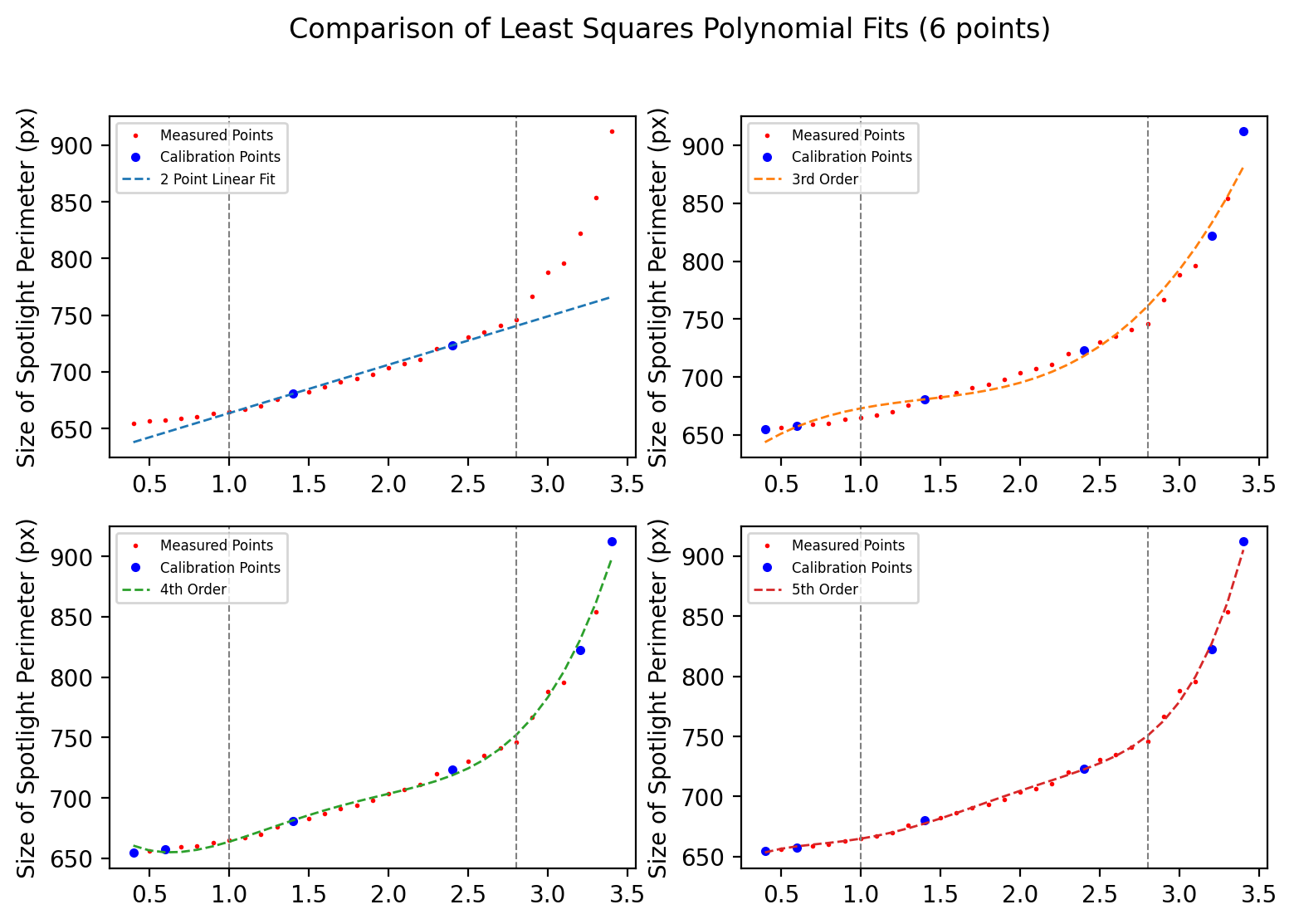}
    \caption{\textbf{Curve comparisons} Comparing 2 point linear fit with 6 point least square polynomial fits of different orders}
    \label{fig:polynomial-curve-compare-by-order}
\end{figure}

In order to represent the entire curve, we need a high order polynomial fit. Using a least squares approximation method, we can fit polynomials to any subset of points we choose. Considering the phenomena shown in figure \ref{fig:data-with-meniscus}, the curve we are attempting to characterize has 3 distinct regions. Calibration points should be selected to include points from each of these regions. Precision measurements of well volume inside the meniscus inversion region is unlikely to be very accurate, as disruptions in surface tension can cause overflow to occur at different points. Therefore for calibration it should be sufficient to model the linear region occurring at the start of meniscus inversion. This is adequate to detect that the fluid level is greater than the capacity of the well. Since the other 2 regions are linear, we can adequately fit a curve with 2 data points selected from each region. Comparisons of different order curve fits can be seen in figure \ref{fig:polynomial-curve-compare-by-order}



\subsubsection{Generalizability of transfer function}
\begin{figure}[H]
    \centering
    \includegraphics[width=\textwidth]{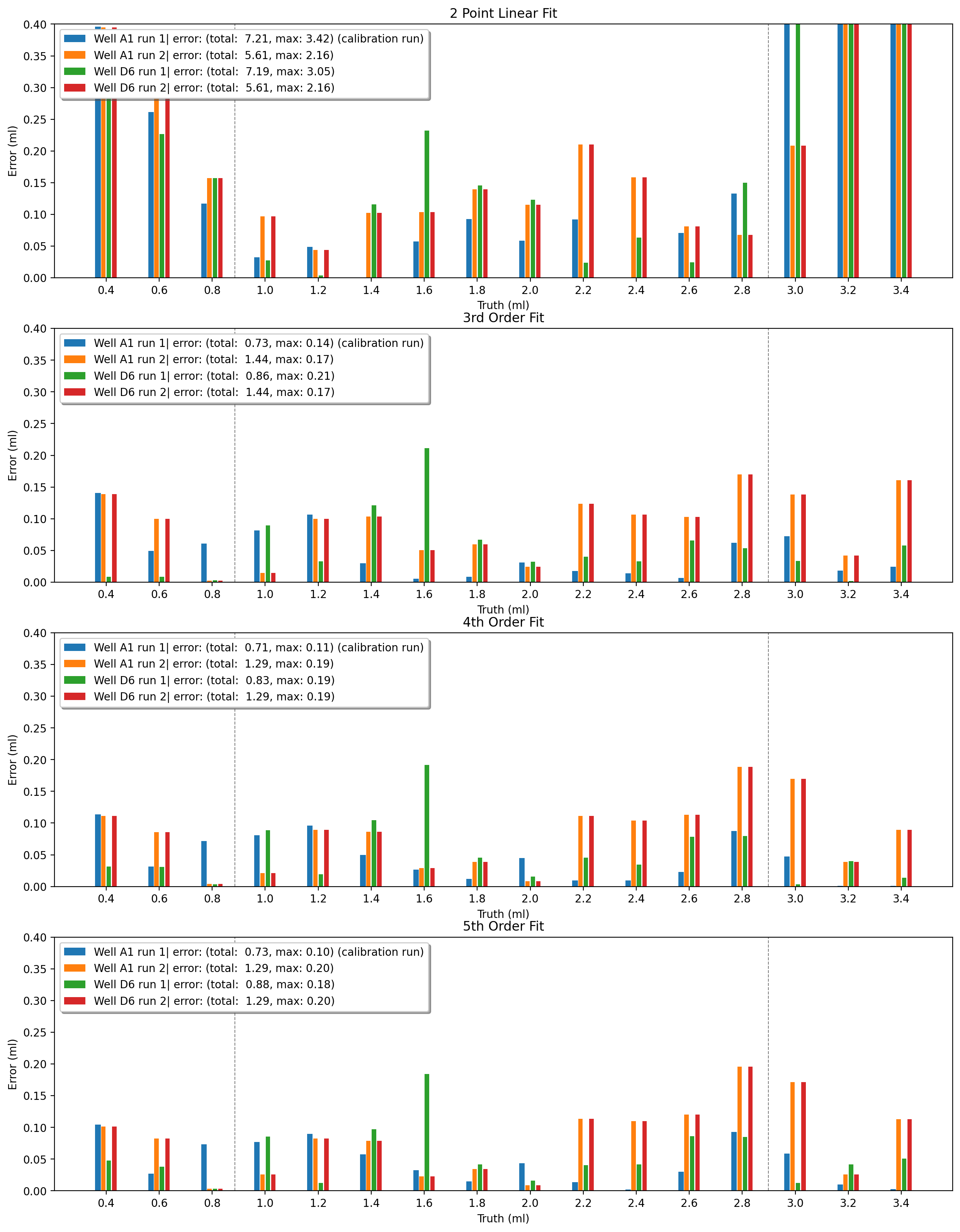}
    \caption{\textbf{Error Comparison} Comparing errors across 4 runs for various polynomial fits}
    \label{fig:polynomial-error-compare-by-order-and-well}
\end{figure}

To evaluate the consistency of this measurement principle, data was captured in 2 different wells during 2 different runs. The resulting error magnitudes are shown in figure \ref{fig:polynomial-error-compare-by-order-and-well}

As expected, the error from the 2 point linear fit is substantial outside the central linear range, but within this range, the measurement error can be reasonable depending on the precision needs of the user. Having the option of a simple 2 point calibration for certain uses is an advantage. The other polynomial fits use the 6 points shown in the curve in figure \ref{fig:polynomial-curve-compare-by-order}. It is interesting to note that the 4th order curve fit performs better than the 5th order fit. Most datapoints shown fall within 100$\mu l$ error with occasional outlier measurements with a maximum error of 190$\mu l$. Knowing this measurement precision means that volumetric fluid dosing measured with this particular implementation of this principle should be tolerant to errors up to 190$\mu l$.



\subsection{Dry Well}
\begin{figure}[H]
    \centering
    \includegraphics[width=0.75\textwidth]{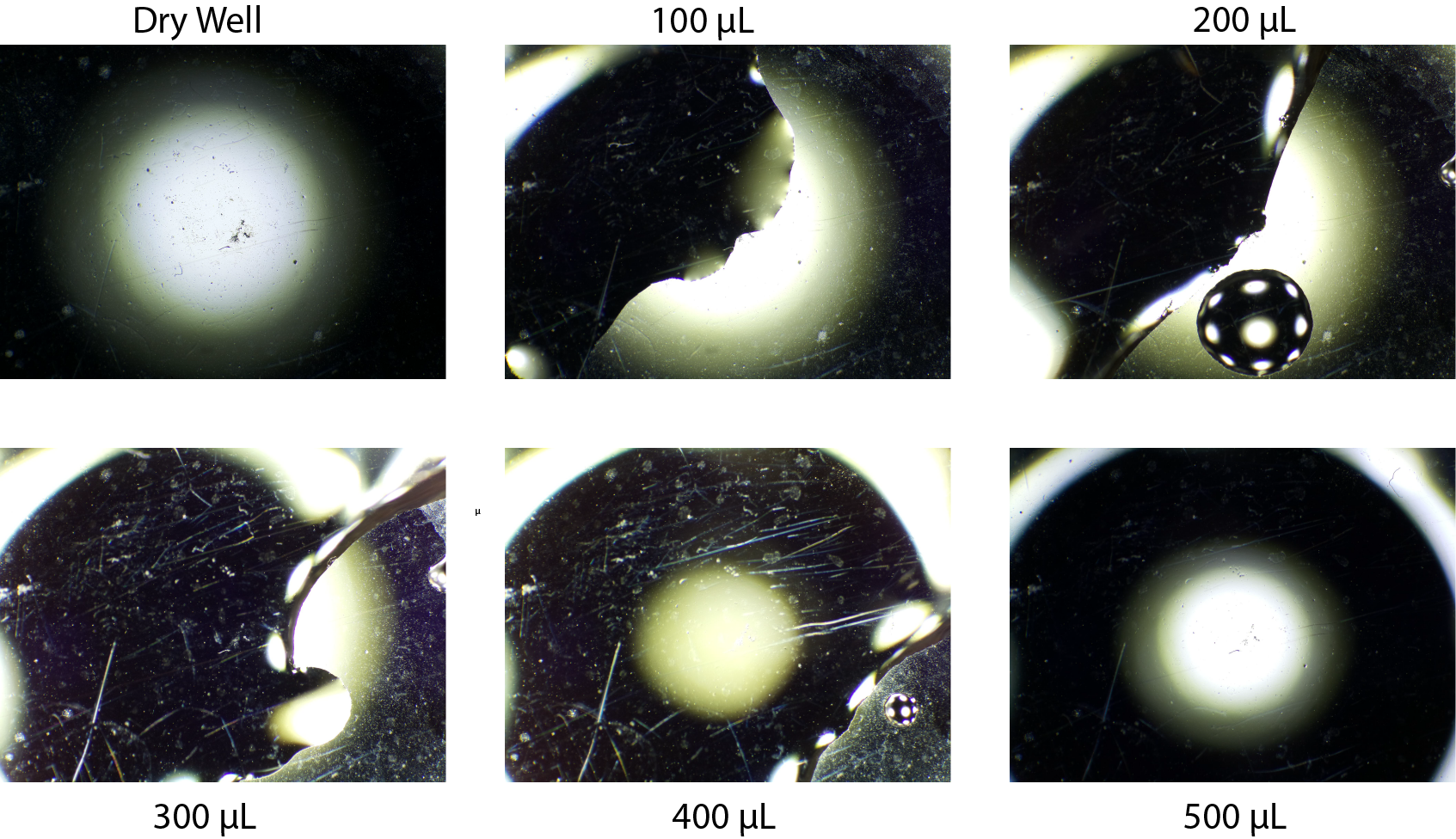}
    \caption{\textbf{Limitations with dry wells} before a fluid film has formed across the dry bottom of the well, measurements have no significance}
    \label{fig:Dry-Fill}
\end{figure}

When a well is completely dry, droplets hold their shape rather than forming a layer along the bottom. Once $~$0.5ml have been placed in the dry well, a layer adheres to the bottom and the measurement starts to work. This is shown in figure \ref{fig:Dry-Fill} The well can also be moistened beforehand with a small amount of water, doing so disrupts the formation of droplets and allows the measurements to work at smaller fluid volumes.

\section{Discussion}\label{sec:discussion}
By leveraging the availability of low cost camera hardware and open source machine vision tools, devices that use computer vision for sensing tasks are likely to become more ubiquitous. The proliferation of cheap mass-produced camera sensors is an opportunity for designers to consider many new methods of measurement.

The device shown here serves as a simple proof of concept for the use of this fluid level measurement principle. For practical purposes, this should be deployed with many cameras in parallel, each monitoring a single well. Previous work has been published laying out the design and use of a 24 well parallel microscope system using similar camera hardware and LEDs, this system is called the "Picroscope" \cite{ly_picroscope_2021, baudin_low_2021}. The  picroscope is also built to be compatible with a microfluidic cell culture feeding platform \cite{seiler_modular_2022}. Applying this principle to this system can enable feedback control for fluid contents of the wells in this and other microfluidic "lab-on-a-chip" type systems.

It would also be possible to use this principle without the overhead LED being permanently affixed above the culture plate. In applications using pipette robots, LEDs could be attached to the end effector, if the arm positions the LED at a known position, spot measurements can be taken and used to detect potential dosing errors or losses due to evaporation.

The principle can also be applied to much larger containers where fluid level measurements are important. Fluid storage containers exist for many applications and while approaches for measuring level in large containers exist, few of those applications are applicable to small volume containers like culture plates. The approach presented here may be broadly applicable to many fields. Furthermore, with different light frequency selections, this principle could be applied to fluids that are opaque in the visible light spectrum but transparent to infrared.

This proposed fluid level sensing method is a simple, reliable, and accurate approach that has been validated for cell culture plates and has potential for uses in other applications.

\begin{backmatter}
\bmsection{Funding}
This work was supported by the Schmidt Futures SF 857, the National Human Genome Research Institute under Award number 1RM1HG011543, the National Science Foundation under award numbers NSF 2034037 and NSF 2134955.

\bmsection{Acknowledgments}
P.V. Baudin thanks David Haussler for the support and guidance he provides to the many projects in our research group.

\bmsection{Disclosures}

\medskip

\noindent PVB: UC Santa Cruz (P)

\medskip

\bmsection{Data Availability}
The data used to generate the plots shown in this article are available from the corresponding author upon request. Code used to measure contour size from live video stream is available from the corresponding author upon request. 3D Model files used for printing the measurement device are available from the corresponding author upon request. This work is intended to be released open source, a GitHub repository containing the measurement code and 3D model files will be made public at time of publication. While repository remains private, access will be granted by request.

\bmsection{Supplemental document}
See Supplement 1 for supporting content. 

\end{backmatter}

\bibliography{sample}






\end{document}


\maketitle

\section{Derivation of Image Distance function}
Labeled values are in reference to items in figure 1 of the main document
\label{supp:derivation}
$$
h = h_0 + L + h_1 
$$
since $h_0$ varies with $L$ but $h$ is constant we reframe $h_0$ as
$$
h_0 = (h - h_1) - L
$$
$h - h_1$ is constant and we will now label it as $C_0$
$$
\begin{aligned}
&x_{0}=(C_0 - L) \tan \theta_{1} \\
&C_1 = \tan \theta_{1} \\
&x_{1}= x_0 + L \tan \theta_{2} \\
&C_2 = \tan \theta_{2} \\
&x_f = x_1 + h_1 C_1 \\
&x_f = (C_0 - L)C_1 + L C_2 +h_1 C_1 \\
&x_f = C_0C_1 - LC_1 + L C_2 +h_1 C_1 \\
&x_f = L(-C_1 + C_2) + h_1C_1 + C_0C_1 \\
& \tan(\theta_1) = \frac{x_f}{I} \\
&I=\frac{x_f}{C_0} \\
&I=\frac{L(-C_1 + C_2) + h_1C_1 + C_0C_1}{C_0}
\end{aligned}
$$